# Machine learning-enabled tomographic imaging of chemical short-range atomic ordering

Yue Li [1]*, Timoteo Colnaghi [2], Yilun Gong [1,7]*, Huaide Zhang [3], Yuan Yu [3], Ye Wei [4], Bin Gan [5], Min Song [5], Andreas Marek [2], Markus Rampp [2], Siyuan Zhang [1], Zongrui Pei [6], Matthias Wuttig [3], Jörg Neugebauer [1], Zhangwei Wang [5]*, Baptiste Gault [1,8]*

[1] Max-Planck Institut für Eisenforschung GmbH, Max-Planck-Straße 1, Düsseldorf, 40237, Germany

[2] Max Planck Computing and Data Facility, Gießenbachstraße 2, Garching, 85748, Germany

[3] Institute of Physics (IA), RWTH Aachen University, Aachen, 52056, Germany

[4] Ecole Polytechnique Fédérale de Lausanne, School of engineering, Rte Cantonale, 1015 Lausanne, Switzerland

[5] State Key Laboratory of Powder Metallurgy, Central South University, Changsha, 410083, China

[6] New York University, New York, NY10012, United States

[7] Department of Materials, University of Oxford, Parks Road, Oxford OX1 3PH, UK

[8] Department of Materials, Imperial College, South Kensington, London SW7 2AZ, UK

*Corresponding authors, yue.li@mpie.de (Y. L.); y.gong@mpie.de (Y. G.); z.wang@csu.edu.cn (Z. W.); b.gault@mpie.de (B. G.)

**In solids, chemical short-range order (CSRO) refers to the self-organisation of atoms of certain species occupying specific crystal sites. CSRO is increasingly being envisaged as a lever to tailor the mechanical and functional properties of materials. Yet quantitative relationships between properties and the morphology, number density, and atomic configurations of CSRO domains remain elusive. Herein, we showcase how machine learning-enhanced atom probe tomography (APT) can mine the near-atomically resolved APT data and jointly exploit the technique's high elemental sensitivity to provide a 3D quantitative analysis of CSRO in a CoCrNi medium-entropy alloy. We reveal multiple CSRO configurations, with their formation supported by state-of-the-art Monte-Carlo simulations. Quantitative analysis of these CSROs allows us to establish relationships between processing parameters and physical properties. The unambiguous characterization of CSRO will help refine strategies for designing advanced materials by manipulating atomic-scale architectures.**

**One-Sentence Summary:** 3D atomic-level details of chemical short-range order are revealed in concentrated alloys by machine learning enhanced atom probe tomography.

Since the Palaeolithic Age, humanity has developed empirical strategies to tailor the properties of materials by manipulating their compositions, structures, and imperfections from the macro- to microscale and even atomic scale. Alloy making traditionally involves the introduction of small quantities of one or more species, solutes, into a matrix of a solvent element. During processing, one or more (meta)stable phases form that modifies the response to physical or mechanical stimulation (*1-4*). In the past decade, so-called high/medium-entropy alloys (H/MEAs) have been introduced, whereby multiple elements are mixed in equal, or close to equal quantity. Although initially assumed to be chemically disordered (*5-9*), i.e., atoms from these principal elements randomly occupy sites of the crystalline lattice, recent studies have suggested that atomic-scale, chemical short-range order (CSRO) is far more prevalent in H/MEAs than initially assumed, offering a potential lever to tailor their properties (*6, 10-16*).

A representative H/MEA is CoCrNi, in which the presence and nature of CSRO are currently debated (*6, 11, 13-18*). Transmission electron microscopy (TEM)-based approaches are most prevalently used to resolve CSRO (*6, 11, 13, 16, 19, 20*), but reports on the presence and configuration of CSRO have been thus far contradictory, even for samples synthesized in the same conditions (Table S1 and Supplementary Text). Due to the intrinsic limit of two-dimensional projection imaging, it has been pointed out that the observed electron reflections may originate from factors other than CSROs, e.g., planar defects and surface oxides (*19, 20*). An alternative, reliable, three-dimensional (3D) analytic perspective of CSRO is hence needed to reconcile these controversies, but also to facilitate the use of CSRO in materials design.

Atom probe tomography (APT) has long been expected to probe CSRO in 3D, but recognising CSRO has been hindered by its anisotropic spatial resolution and imperfect detection efficiency (*21-23*). Overcoming these limitations by manual analysis has proven impossible (*24-26*). Inspired by machine learning (ML) deployed to process complex microscopy and microanalysis data (*27-29*), and building on our previous efforts (*23, 30*), we introduce a bottom-up approach to quantify in 3D the CSRO domains in APT data from CoCrNi, termed ML-APT, that does not require any prior knowledge of the CSRO configurations. The overall flowchart is presented in Fig. S1. ML-APT enables the identification of CRSOs as well as the quantification of the number density of ordered domains, their configurations, elemental site occupancy, and size/morphology. Monte-Carlo simulations are then used to rationalise our analyses, facilitating an understanding of ordering reactions. We finally showcase how to establish a direct processing-CSRO-property relationship, paving the way for further material design opportunities.

**Results and discussion**
*APT results*
Two equiatomic CoCrNi alloys were analysed following homogenisation, and after homogenisation and annealing (Methods, Table S2). We performed correlative scanning electron microscopy (SEM)-electron backscattered diffraction (EBSD)-focused ion beam (FIB)-APT to characterise their microstructure in grains of selected orientation (Fig. 1A), i.e. {002} and {111}. Fig. 1B-E detail an APT analysis from the annealed sample. Fig. 1B is a detector hit map with a pattern corresponding to the symmetries of the {002} crystallographic planes, and Fig. 1C is the 3D atom map reconstructed around this pole. A close-up in Fig. 1D shows resolved {002} atomic planes. The reconstruction was calibrated to the reported interplanar spacing of face-centred-cubic (FCC) CoCrNi (*31, 32*). Spatial distribution maps (*33, 34*) are calculated along the depth (z-SDMs) to exploit these most-highly resolved signals and evaluate the CSRO. The *z*-SDM indicates the characteristic period of each elemental pair along a specific direction, which is similar to a split pair correlation function used in e.g. TEM (*12*). The z-SDMs of different elemental pairs obtained in a 2-nm voxel are plotted in Fig. 1E. The peak-to-peak distance for each elemental pair is the same, suggesting a homogenous solid solution. The typical clustering algorithms in the APT community (*26, 35, 36*) have been tested but cannot identify CSROs (Fig. 1 F and G, Supplementary Text). A similar analysis along the {111} planes is provided in Fig. S2. The spatial resolution for the {022} planes is insufficient to perform the subsequent analyses.

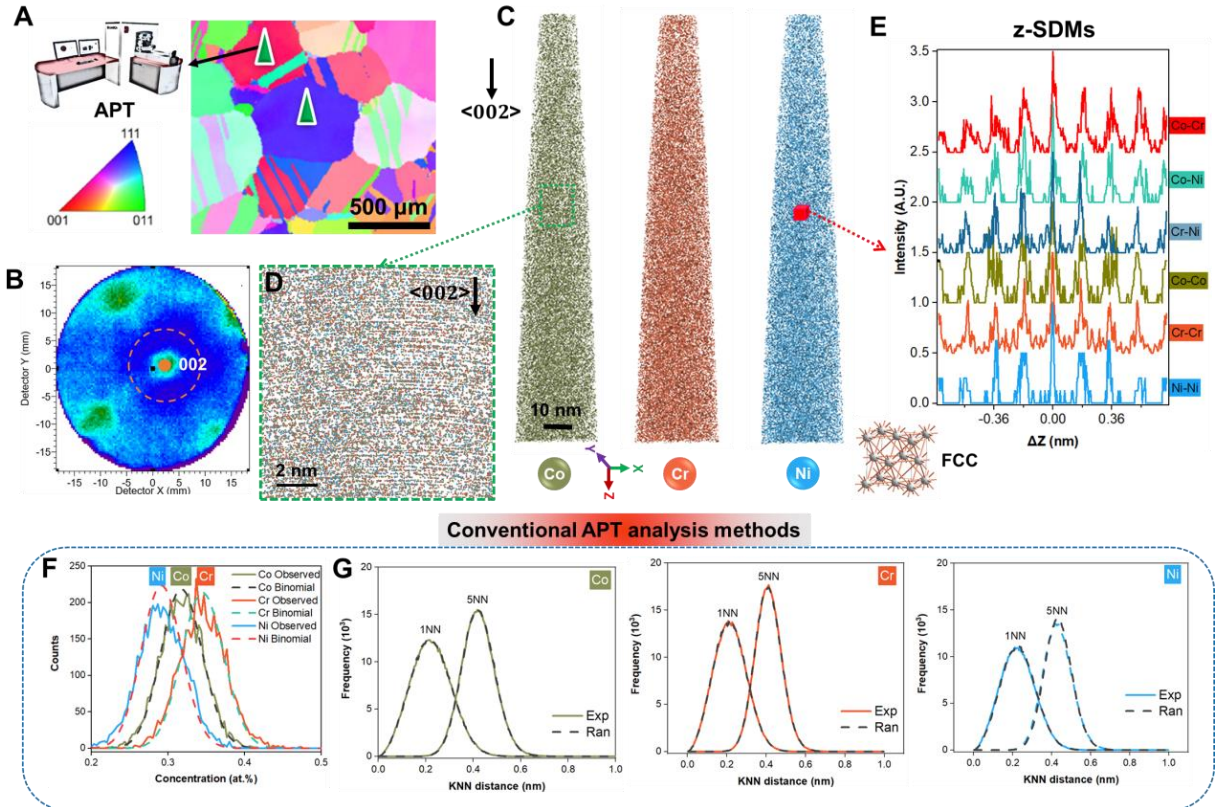

**Fig. 1 APT data of a typical equiatomic CoCrNi alloy annealing at 1273K for 120h and conventional data analysis to look for CSRO.** (A) The EBSD inverse pole figure highlights the grains used for APT experiments. (B) Representative 2D detector hit map. One centric {002} crystallographic pole is labelled. (C) Precise 3D APT reconstruction along the <002> orientation. (D) Local close-up of a thin slice in (C) along the <002>. (E) z-SDMs of different elemental pairs in a representative 2-nm voxel in (C). Its signature corresponds to the FCC structure and its unit cell is given. Two kinds of conventional APT analysis approaches: (F) frequency distribution analysis of Co, Cr, and Ni atoms compared to the binomial random distributions, and (G) K-nearest neighbour (KNN) distance analysis (K=1 and 5) of Co-Co, Cr-Cr, and Ni-Ni elemental pairs. Exp and Ran labels correspond to the results obtained by experimental and random-labelled datasets, respectively.

*ML-APT framework*

As detailed in Fig. 2A for $L1_2$-CSRO, for the random solid solution of FCC-based CoCrNi alloys, the elemental occupation of each site is equiprobable. CSRO occurs when particular sites have a higher probability to be occupied by a specific element, e.g. the face-centred sites are more likely to be Cr/Ni while the edges are Co atoms. At higher probability, up to close to 100%, CSRO is established and can facilitate the nucleation of long-range chemical order. The corresponding Co-Co z-SDMs along the <002> from simulated APT data are shown in Fig. 2B, and the peaks close to ±0.18nm and ±0.54nm are disappearing with the evolution of CSRO. Any kind of CSRO can be detected, provided that its signature in the z-SDMs along a particular orientation is clear. This allows us to recognise different CSRO configurations without any prior knowledge, which is unlike the previous up-bottom strategy with prior possible ordered or CSRO structures (*23, 30*).

The ML-APT workflow to reveal CSRO in H/MEA is as follows. First, we generated artificial APT data along the <002> or <111> containing either a randomly distributed FCC-matrix or CSRO (Methods). Over 10000 of the corresponding z-SDMs patterns are recorded for each orientation (Table S3). This synthetic data is fed into an optimized 1D convolutional neural network (CNN) to obtain an FCC-matrix/CSRO binary classification model (Fig. 2C, Fig. S3A). ML-APT shows excellent performance for both simulated and experimental test datasets (Fig. S3 B-D, Fig. S4). It was further tested on large-scale CoCrNi artificial APT data with $L1_2$-CSRO domains with a diameter of 0.7–2nm, and ML-APT distinguishes these well (Fig. S5, Supplementary Text). Gradient-weighted class activation mapping (*23, 37*), which uses the gradients of any target concept flowing into the final convolutional layer to produce a coarse localization map highlighting the important regions in the image for predicting the concept, reveals that ML-APT performs the classification by focusing on the specific peaks of the z-SDMs that can be used to accurately classify the FCC/CSRO (Fig. S6). Finally, experimental z-SDMs were subjected to pre-processing and then input into ML-APT to obtain the 3D CSRO distributions (Fig. 2D, Methods).

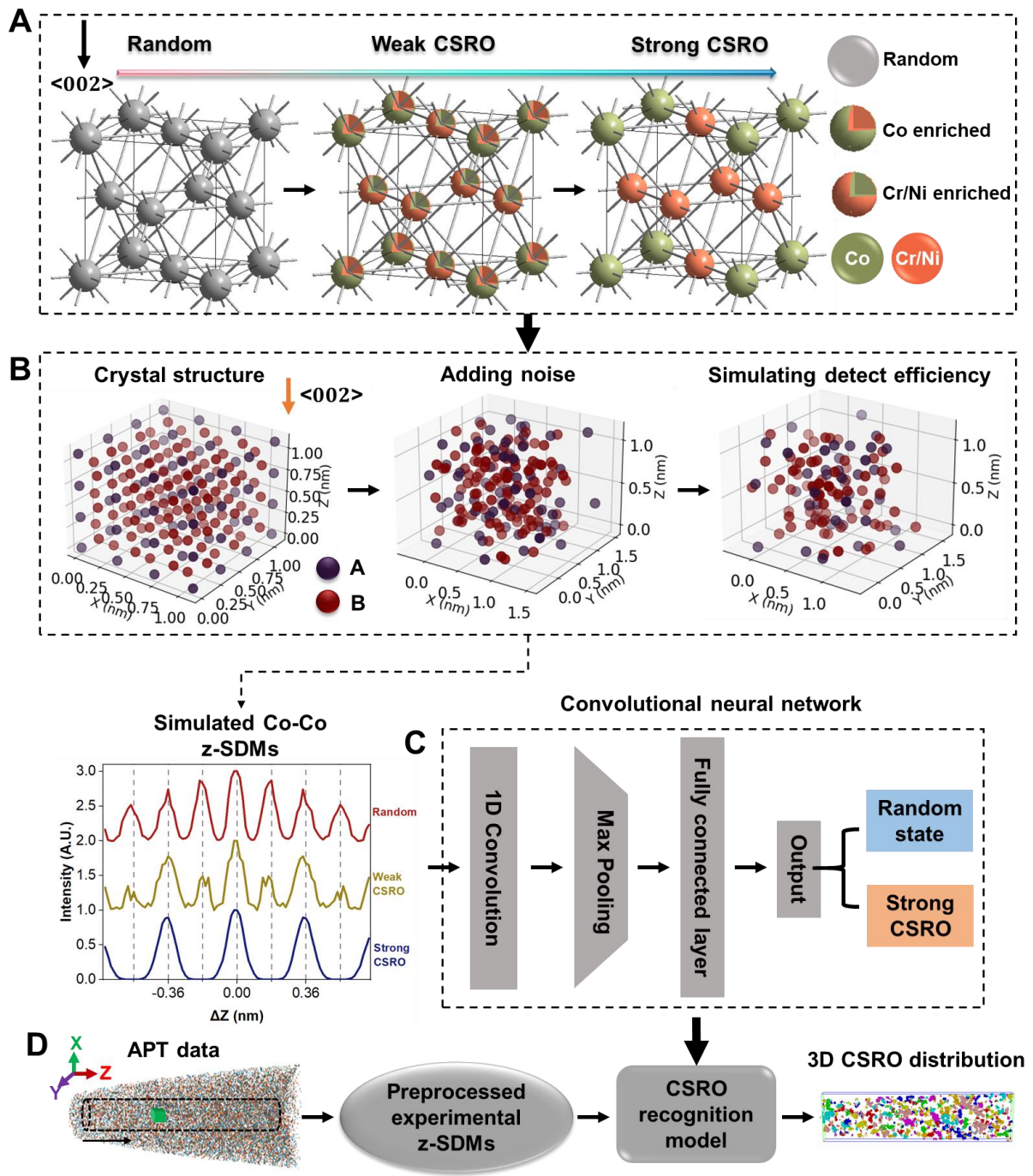

**Fig. 2 Proposed ML-APT framework to recognise multi-type CSROs in CoCrNi alloys.** (A) Unit cells of random-FCC, weak and strong $L1_2$-CSRO. (B) Typical Co-Co *z*-SDMs along the <002> with the evolution of CSRO after performing APT simulation. (C) Schematic diagram of the optimised 1D CNN structure to obtain a random-FCC/CSRO recognition model. (D) Flowchart of processing experimental data to obtain 3D CSRO distribution.

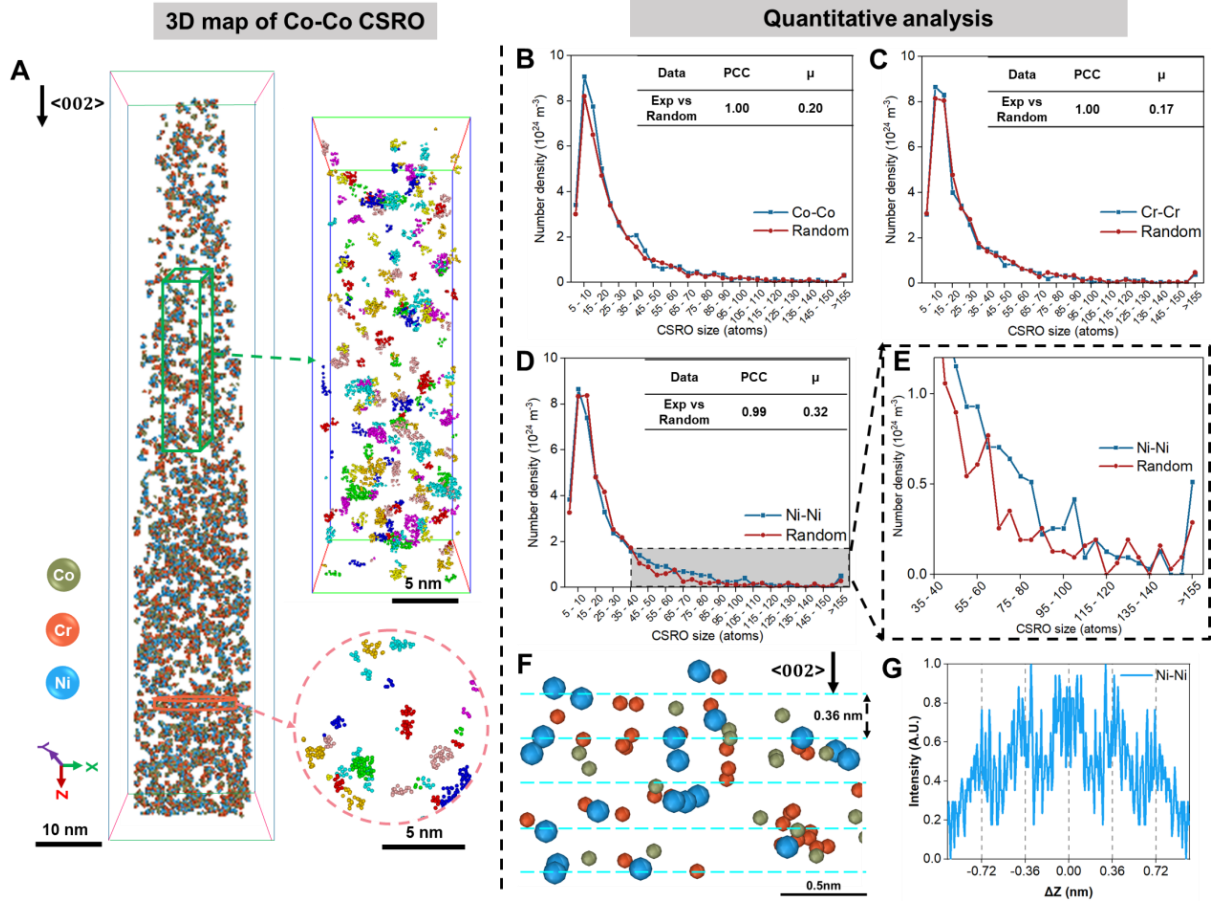

**Fig. 3 3D quantitative analysis of CSRO along <002> in the annealed CoCrNi alloy.** (A) 3D distribution of Co-Co CSROs with the mapping of elements. The front and top views of typical clusters are provided. Different colours mark different CSRO domains. (B), (C), (D) Size distributions of the identified Co-Co, Cr-Cr, and Ni-Ni CSROs, respectively. The results from the chemically-randomised dataset (Methods) are compared with the PCC and Pearson contingency coefficient ($\mu$). This CSRO size refers to the APT-counted atoms and the size of 55 atoms ideally corresponds to a 1-nm cube. (E) Local enlargement of the coloured region in (D) which is different from the random curve. (F) 3D atom map of a typical Ni-Ni CSRO domain extracted from (E). (G) Its corresponding Ni-Ni z-SDM.

*3D perspective of CSRO*
Typical examples of the 3D distributions of CSRO domains obtained from ML-APT, applied to Co-Co, Cr-Cr, and Ni-Ni, are presented in Fig. 3 and Fig. S7 along the <002> and <111>, respectively. The cross-species elemental pairs were not analysed to avoid possible biases arising from differences in evaporation fields affecting the spatial resolution. Fig. 3A shows a homogenous distribution of these domains with a near-spherical morphology (Fig. S8). Fig. 3B–D shows the size distributions of domains in which the Co-Co, Cr-Cr, and Ni-Ni are classified as ordered, respectively. Pearson's correlation coefficients (PCC) and contingency coefficient ($\mu$) (*38*) are used to test the statistical significance of the difference between these distributions and a chemically-randomized dataset, with $\mu$ found more sensitive than PCC to characterise such subtle differences. We defined a threshold to classify (non-)randomness at 0.25. The Ni-Ni distribution is non-random, with a $\mu$ of 0.32, especially when the domains have more than 35 atoms (<1nm) (Fig. 3E), while the distribution of the two other elements is closer to random ($\mu$ <0.25). Fig. 3F is an example of the Ni-Ni CSRO domain, and the corresponding Ni-Ni $z$-SDM is plotted in Fig. 3G, showcasing an interplanar spacing of Ni atoms is twice as large as that in the FCC-matrix (Fig. S1E), which matches the $L1_2/DO_{22}$-type structure with the Ni-Ni repulsion on {100} as explained in Fig. 4A. For comparison, along <111>, the three kinds of CSROs are all different from the random state with $\mu \geq 0.25$ (Fig. S7, B-D), which matches the $L1_1$-type structure with the Co/Cr/Ni repulsion on {111}, as detailed in Fig. 4B.

Fig. 4 C and D provide values of $\mu$ for the two studied material states and orientations. Along the {002} planes, the $\mu$ of Co-Co or Cr-Cr CSROs remains below 0.25. Non-statistical Ni-Ni CSRO rises from 0.18 to 0.27 after annealing at 1273K for 120h, with a number density of $3.88 \times 10^{24}$ m$^{-3}$ of CSRO domains with atomic configurations matching the $L1_2/DO_{22}$ structures with the Ni-Ni repulsion on {100} (Fig. 4 A and E). Note that the probability for the $L1_0$ structure is much lower compared to $L1_2/DO_{22}$. Along {111} planes, after homogenization, values of $\mu$ for Co-Co and Ni-Ni CSRO are close to or above 0.25, suggesting the existence of $L1_1$-domains with the Co/Ni repulsion on {111}, with a number density in the range of $2.24 - 3.70 \times 10^{24}$ m$^{-3}$ (Fig. 4 B and E). After annealing, the $\mu$ of Co-Co, Cr-Cr, and Ni-Ni pairs are above 0.25, matching with the $L1_1$-type structure with the Co/Cr/Ni repulsion on {111}, with a number density of $4.35 \times 10^{24}$ m$^{-3}$, $3.59 \times 10^{24}$ m$^{-3}$, and $4.75 \times 10^{24}$ m$^{-3}$, as determined from Co, Cr and Ni pairs, respectively (Fig. 4 B and E). Only $L1_1$-domains exist after homogenization, and their number density increases after annealing, during which a high density of $L1_2/DO_{22}$-domains appear (Fig. 4 E and F). Overall, the number density of multiple CSROs is approx. 2.8 times after annealing compared to after homogenization. During the entire process, the CSROs almost keep spherical shape (Fig. S8) with the size of 20~155 APT-counted atoms (0.7-1.5nm in diameter) (Fig. 3 and Fig. S7). Note that these observed domains along {002} and {111} are not the same ones.

*Electrical response*
The occurrence of CSRO in solid solutions is often associated with the modifications of physical properties (*24, 25, 39, 40*). The influence of CSRO on the mechanical properties of CoCrNi has been studied widely (*6, 13, 19, 31*), with inconsistent conclusions, but functional properties have only rarely been investigated. Here, we measured the electrical resistivity of the two material states (Methods). The annealing-induced multiple CSROs in CoCrNi alloys resulted in a 17% rise in the room-temperature electrical resistivity (Fig. 4G), higher than previous reports (+4.8%) in Ref. (*19*), which can be explained by the formation of a higher density of CSRO domains during furnace cooling compared to a quench. This reveals a high sensitivity of the electrical response upon changes in the CSRO state, maybe more so than the mechanical response. This remarkable increase in resistivity implies that the increasing CSRO might lead to a reduced electronic density of states at the Fermi level, consistent with the previous density functional theory calculations in the CoCrNi system (*41*).

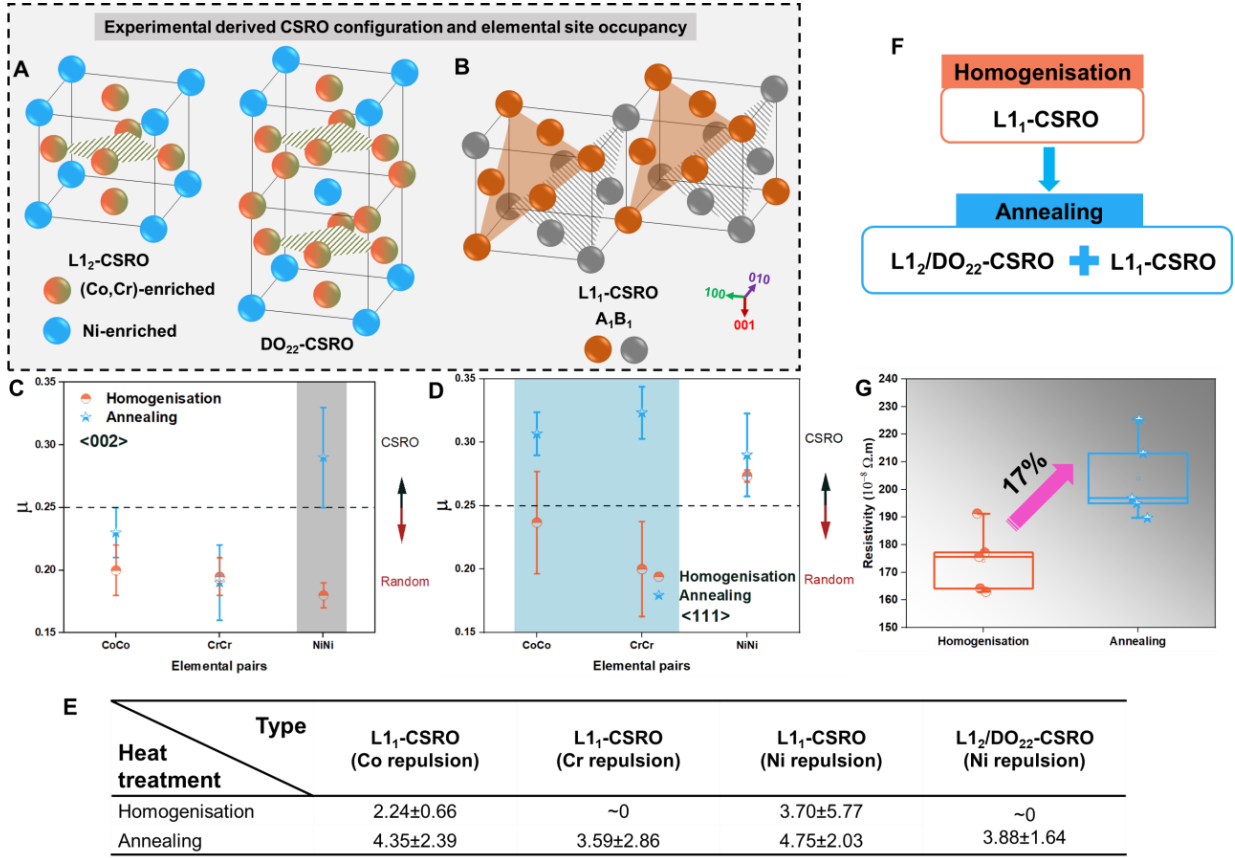

**Fig. 4 3D atomic-level details of multi-type CSROs in CoCrNi alloys under different heat treatments and arising electrical resistivity change.** (A) The $L1_2/DO_{22}$-CSRO structure with the Ni-Ni repulsion on {100}. (B) The $L1_1$-CSRO structure with the A-A or B-B repulsion on {111}. Element A or B refers to any Co, Cr, or Ni atom but cannot be the same simultaneously. (C) and (D) Changes of Pearson contingency coefficient ($\mu$) under different heat treatments along <002> and <111>, respectively. The coloured regions highlight the changes of $\mu$ after annealing. Three APT datasets were analysed to obtain the statistical results for each data point. A value of $\mu=0.25$ is regarded as the threshold between CSRO and random states. (E) Number-density change ($10^{24}$ m$^{-3}$) of different types of CSROs under heat treatment. The number density is obtained after subtracting the random-state value. (F) Derived CSRO structural evolution from homogenisation to annealing. The corresponding CSRO configurations are plotted in (A) and (B). (G) Evolutions of electrical resistivity under different heat treatments.

*Monte-Carlo simulation*

ML accelerated *ab-initio* Monte-Carlo (MC) simulations (Methods) were performed to predict the temperature-dependent equilibrium CSROs and associated crystalline structures. Predicted temperature-dependent heat capacities (relevant to ground state total energy and configurational entropy) suggested two peaks due to first-order phase transformations (Fig. S9). One occurs at around 860K, which is confirmed by differential scanning calorimetry (DSC), and the second is at around 200K which is below the detection limit of DSC due to sluggish diffusion kinetics at low temperatures. Predicted 1$^{st}$ NN Warren-Cowley parameters (Methods) suggested repulsion of Cr-Cr, Co-Ni, Co-Co and attractions of Ni-Ni, Cr-Ni, Co-Cr above the phase transformation peak, as shown in Fig. S10. To identify the crystalline structure of CSROs being predicted, Fig. 5 A and B visualize the calculated CSRO diffuse intensity map ($\alpha_\mathbf{q}$) (Methods) in the (001) and (111) planes, respectively, at 1000K. For the (001), a (1, 0.5, 0) special point is revealed instead for Cr-Cr, suggesting its $DO_{22}$ ordering (*42*). Besides, a strong (001) peak is also presented for the Ni-Ni and Co-Co pairs: $L1_2$ or $L1_0$ structure is suggested for these two. Further peak analysis in the 3D reciprocal space found (0.5, 0.5, 0.5) maxima for Cr-Cr and Ni-Ni pairs, which are also revealed by projecting the calculated $\alpha_\mathbf{q}$ in the (111) as shown in Fig. 5B, suggesting their coexistence with the $L1_1$ structure (i.e. ordering along <111>). Compared with APT measurements (Fig. 4), ordering along <002> for Ni-Ni, as well as ordering along <111> for Cr-Cr and Ni-Ni are confirmed by MC simulations. However, no significant $L1_1$ ordering is predicted for the Co-Co pair.

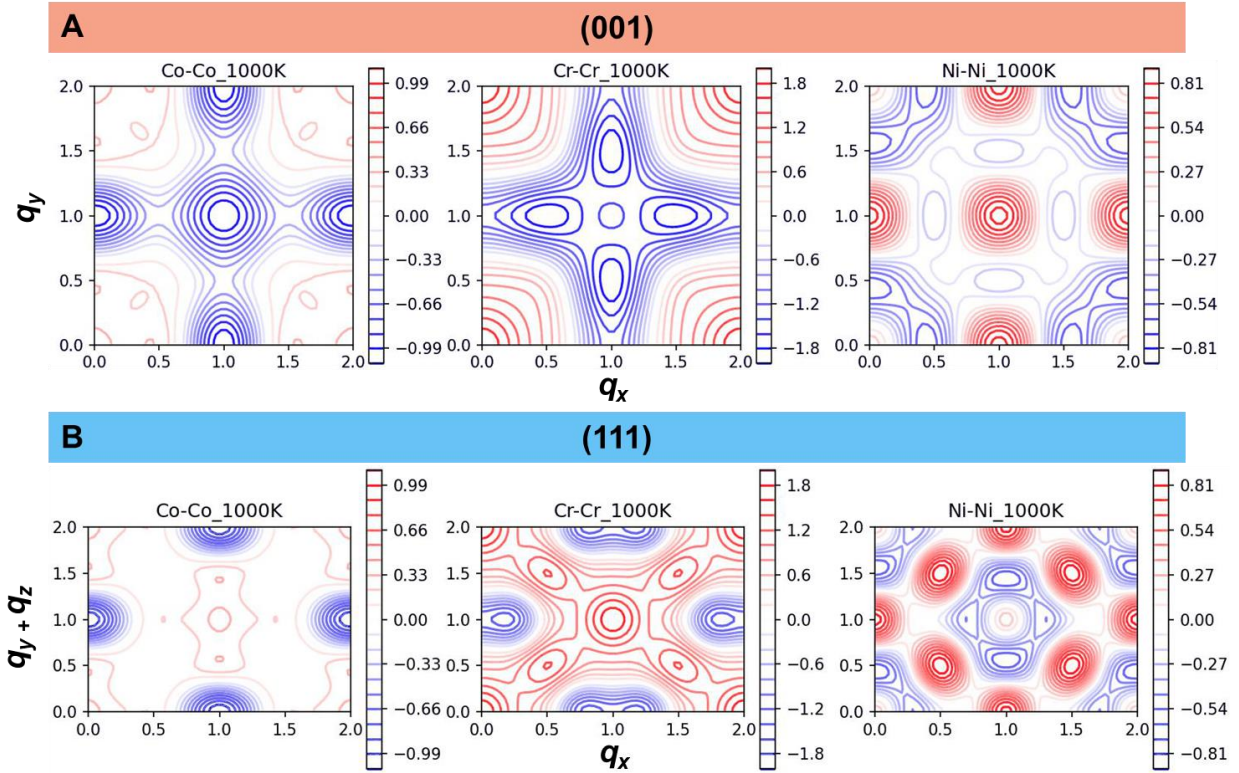

**Fig. 5 CSROs predicted by Monte-Carlo simulations.** (A) and (B) Predicted CSRO diffuse intensity map, $\alpha_\mathbf{q}$, of Co-Co, Cr-Cr and Ni-Ni pairs at 1000K in (001) and (111) planes, respectively; reciprocal space vectors are given in units of $2\pi/a$.

## Conclusions

To conclude, the proposed ML-APT approach enables us to settle previous debates on CSRO in CoCrNi alloys, and evidence atomic-scale details of CSRO beyond the state-of-the-art. In comparison to others approaches for CSRO characterisation, our proposed ML-APT analysis only relies on the measured APT mass spectra and is hence independent on interatomic potentials needed to interpret CSRO from X-ray/neutron techniques (*17, 18*). Moreover, ML-APT provides 3D elemental-specific information and is hence capable of identifying multiple CSROs (Fig. 3 and Fig. 4). Most TEM-based techniques only have access to 2D information integrated over an ensemble of 3D CSRO regions (*6, 11, 13, 16, 19, 20*), making it challenging to detect small domains or separate CSRO from possible artefacts originating from e.g., surface steps, planar defects, and oxide layers (*19, 20*). The observed CSRO configurations were here rationalized herein by Monte-Carlo simulations. The electrical response is found to be a better indicator of CSROs than mechanical properties. The individual influence of CSRO on the mechanical properties is seemly limited at room temperature, whereas, the co-existing CSRO and medium-range order could be a better approach to designing new H/MEAs with better (cryogenic) mechanical properties via adjusting the processing parameters (e.g., thermal history (*43*) and deformation (*44, 45*)) or microalloying (e.g., adding Ti or Al (*46, 47*)). Our method can be generally used for other H/MEAs as well as for complex engineering materials. Advancing the understanding of CSRO reactions will help with integrating them into the design of future high-performance alloys.

## Acknowledgements

This work was primarily supported by the research fellowship provided by the Alexander von Humboldt Foundation and the Max Planck research network on big-data-driven materials science (BiGmax). BG acknowledges financial support from the ERC via the funding of the project SHINE 771602. YG acknowledges the initial implementation of LRP and MC/Python simulation packages by Alexander Shapeev, and funding from Deutsche Forschungsgemeinschaft (SPP 2006) and Next Generation TATARA Project sponsored by the Cabinet of Japan. ZW acknowledges Natural Science Foundation of Hunan Province (Grant No. 2022JJ30712). Frank Stein and Andreas Jansen are acknowledged for their help with DSC measurements. Uwe Tezins and Andreas Sturm are acknowledged for their support of the FIB and APT facilities. Alisson Kwiatkowski da Silva is acknowledged for fruitful discussions.

## Author contributions

YL, YG, ZW, and BG designed the project. YL is the lead experimental/data scientist. MS and BiG prepared the materials and heat treatments. YL performed the FIB/APT experiments and analysed the APT data with the help of BG. YL programmed the machine learning framework with the help of YW and TC. HZ and YY performed the electrical resistivity measurements. YG programmed and performed the MC simulations. YL, YG, ZW, and BG wrote the manuscript. All authors contributed to the discussion of the results and commented on the manuscript.

## Data availability

All data that support the findings are involved in this paper. Other data are available from the corresponding authors upon reasonable request.

## Code availability
The ML-APT software is available at the GitHub address https://github.com/a356617605. Other codes are available from the corresponding authors upon reasonable request.

## Competing interests
The authors declare that they have no competing interests.

## Supplementary materials:
Methods
Supplementary Text
Figs. S1 to S13
Tables S1 to S3
References

## References and Notes

1. W. Sun *et al.*, Precipitation strengthening of aluminum alloys by room-temperature cyclic plasticity. *Science* **363**, 972-975 (2019).
2. Y. Li *et al.*, Precipitation and strengthening modeling for disk-shaped particles in aluminum alloys: size distribution considered. *Materialia* **4**, 431-443 (2018).
3. J. Zhang *et al.*, Deformation-induced concurrent formation of 9R phase and twins in a nanograined aluminum alloy. *Acta Mater.*, 118540 (2022).
4. R. Zhang *et al.*, Direct imaging of short-range order and its impact on deformation in Ti-6Al. *Sci. Adv.* **5**, eaax2799 (2019).
5. E. P. George, D. Raabe, R. O. Ritchie, High-entropy alloys. *Nat. Rev. Mater.* **4**, 515-534 (2019).
6. R. Zhang *et al.*, Short-range order and its impact on the CrCoNi medium-entropy alloy. *Nature* **581**, 283-287 (2020).
7. C. Niu, C. R. LaRosa, J. Miao, M. J. Mills, M. Ghazisaeidi, Magnetically-driven phase transformation strengthening in high entropy alloys. *Nat. Commun.* **9**, 1363 (2018).
8. F. Zhang *et al.*, Polymorphism in a high-entropy alloy. *Nat. Commun.* **8**, 15687 (2017).
9. S. Zhao, G. M. Stocks, Y. Zhang, Stacking fault energies of face-centered cubic concentrated solid solution alloys. *Acta Mater.* **134**, 334-345 (2017).
10. X. Chen *et al.*, Direct observation of chemical short-range order in a medium-entropy alloy. *Nature* **592**, 712-716 (2021).
11. L. Zhou *et al.*, Atomic-scale evidence of chemical short-range order in CrCoNi medium-entropy alloy. *Acta Mater.* **224**, 117490 (2022).
12. Q. Ding *et al.*, Tuning element distribution, structure and properties by composition in high-entropy alloys. *Nature* **574**, 223-227 (2019).
13. M. Zhang *et al.*, Determination of peak ordering in the CrCoNi medium-entropy alloy via nanoindentation. *Acta Mater.* **241**, 118380 (2022).
14. J.-P. Du *et al.*, Chemical domain structure and its formation kinetics in CrCoNi medium-entropy alloy. *Acta Mater.*, 118314 (2022).
15. F. Walsh, M. Asta, R. O. Ritchie, Magnetically driven short-range order can explain anomalous measurements in CrCoNi. *Proc. Natl. Acad. Sci. U.S.A.* **118**, e2020540118 (2021).
16. H. Hsiao *et al.*, Data-driven electron-diffraction approach reveals local short-range ordering in CrCoNi with ordering effects. *Nat. Commun.* **13**, 6651 (2022).
17. F. X. Zhang *et al.*, Local Structure and Short-Range Order in a NiCoCr Solid Solution Alloy. *Phys. Rev. Lett.* **118**, 205501 (2017).
18. K. Inoue, S. Yoshida, N. Tsuji, Direct observation of local chemical ordering in a few nanometer range in CoCrNi medium-entropy alloy by atom probe tomography and its impact on mechanical properties. *Phys. Rev. Mater.* **5**, 085007 (2021).

19. L. Li *et al.*, Evolution of short-range order and its effects on the plastic deformation behavior of single crystals of the equiatomic Cr-Co-Ni medium-entropy alloy. *Acta Mater.* **243**, 118537 (2023).
20. F. Walsh, M. Zhang, R. O. Ritchie, A. M. Minor, M. Asta, Extra electron reflections in concentrated alloys may originate from planar defects, not short-range order. *arXiv preprint arXiv:2210.01277*, (2022).
21. B. Gault *et al.*, Atom probe tomography. *Nat. Rev. Methods Primers* **1**, 51 (2021).
22. B. Gault *et al.*, Reflections on the spatial performance of atom probe tomography in the analysis of atomic neighbourhoods. *Microsc. Microanal.*, 1-11 (2021).
23. Y. Li *et al.*, Convolutional neural network-assisted recognition of nanoscale L12 ordered structures in face-centred cubic alloys. *npj Comput. Mater.* **7**, 8 (2021).
24. R. K. W. Marceau *et al.*, Atom probe tomography investigation of heterogeneous short-range ordering in the 'komplex' phase state (K-state) of Fe–18Al (at.%). *Intermetallics* **64**, 23-31 (2015).
25. R. K. W. Marceau, A. V. Ceguerra, A. J. Breen, D. Raabe, S. P. Ringer, Quantitative chemical-structure evaluation using atom probe tomography: Short-range order analysis of Fe–Al. *Ultramicroscopy* **157**, 12-20 (2015).
26. R. Hu, S. Jin, G. Sha, Application of atom probe tomography in understanding high entropy alloys: 3D local chemical compositions in atomic scale analysis. *Prog. Mater Sci.* **123**, 100854 (2022).
27. F. Oviedo *et al.*, Fast and interpretable classification of small X-ray diffraction datasets using data augmentation and deep neural networks. *npj Comput. Mater.* **5**, 60 (2019).
28. J. A. Aguiar, M. L. Gong, R. R. Unocic, T. Tasdizen, B. D. Miller, Decoding crystallography from high-resolution electron imaging and diffraction datasets with deep learning. *Sci. Adv.* **5**, eaaw1949 (2019).
29. Y.-F. Shen, R. Pokharel, T. J. Nizolek, A. Kumar, T. Lookman, Convolutional neural network-based method for real-time orientation indexing of measured electron backscatter diffraction patterns. *Acta Mater.* **170**, 118-131 (2019).
30. Y. Li *et al.*, Quantitative three-dimensional imaging of chemical short-range order via machine learning enhanced atom probe tomography. *Preprint*, (2022).
31. B. Yin, S. Yoshida, N. Tsuji, W. Curtin, Yield strength and misfit volumes of NiCoCr and implications for short-range-order. *Nat. Commun.* **11**, 1-7 (2020).
32. M. P. Agustianingrum, S. Yoshida, N. Tsuji, N. Park, Effect of aluminum addition on solid solution strengthening in CoCrNi medium-entropy alloy. *J. Alloys Compd.* **781**, 866-872 (2019).
33. B. P. Geiser, T. F. Kelly, D. J. Larson, J. Schneir, J. P. Roberts, Spatial Distribution Maps for Atom Probe Tomography. *Microsc. Microanal.* **13**, 437-447 (2007).
34. M. P. Moody, B. Gault, L. T. Stephenson, D. Haley, S. P. Ringer, Qualification of the tomographic reconstruction in atom probe by advanced spatial distribution map techniques. *Ultramicroscopy* **109**, 815-824 (2009).
35. P. Dumitraschkewitz, S. S. A. Gerstl, L. T. Stephenson, P. J. Uggowitzer, S. Pogatscher, Clustering in Age-Hardenable Aluminum Alloys. *Adv. Eng. Mater.* **20**, 1800255 (2018).
36. B. Gault, M. P. Moody, J. M. Cairney, S. P. Ringer, *Atom probe microscopy*. (Springer, 2012), vol. 160.
37. R. R. Selvaraju *et al.*, Grad-cam: Visual explanations from deep networks via gradient-based localization. *Proc. IEEE Int. Conf. Comput. Vis.*, 618-626 (2017).
38. M. P. Moody, L. T. Stephenson, A. V. Ceguerra, S. P. Ringer, Quantitative binomial distribution analyses of nanoscale like-solute atom clustering and segregation in atom probe tomography data. *Microsc. Res. Tech.* **71**, 542-550 (2008).
39. H. Thomas, Über widerstandslegierungen. *Z. Phys.* **129**, 219-232 (1951).
40. Z. Pei, R. Li, M. C. Gao, G. M. Stocks, Statistics of the NiCoCr medium-entropy alloy: Novel aspects of an old puzzle. *npj Comput. Mater.* **6**, 122 (2020).
41. A. Tamm, A. Aabloo, M. Klintenberg, M. Stocks, A. Caro, Atomic-scale properties of Ni-based FCC ternary, and quaternary alloys. *Acta Mater.* **99**, 307-312 (2015).
42. U. Kulkarni, S. Banerjee, Phase separation during the early stages of ordering in Mi3Mo. *Acta Metall.* **36**, 413-424 (1988).
43. G. Tang *et al.*, Quantifying chemical fluctuations around medium-range orders and its impact on dislocation interactions in equiatomic CrCoNi medium entropy alloy. *Mater. Des.* **225**, 111572 (2023).
44. M. Heczko *et al.*, Role of deformation twinning in fatigue of CrCoNi medium-entropy alloy at room temperature. *Scripta Mater.* **202**, 113985 (2021).
45. Y. M. Eggeler, K. V. Vamsi, T. M. Pollock, Precipitate Shearing, Fault Energies, and Solute Segregation to Planar Faults in Ni-, CoNi-, and Co-Base Superalloys. *Annual Review of Materials Research* **51**, 209-240 (2021).

46. Y. Zhao *et al.*, Heterogeneous precipitation behavior and stacking-fault-mediated deformation in a CoCrNi-based medium-entropy alloy. *Acta Mater.* **138**, 72-82 (2017).
47. N. Yao *et al.*, Ultrastrong and ductile additively manufactured precipitation-hardening medium-entropy alloy at ambient and cryogenic temperatures. *Acta Mater.* **236**, 118142 (2022).
48. B. Gault *et al.*, Estimation of the Reconstruction Parameters for Atom Probe Tomography. *Microsc. Microanal.* **14**, 296-305 (2008).
49. B. Gault *et al.*, Advances in the calibration of atom probe tomographic reconstruction. *J. Appl. Phys.* **105**, 034913 (2009).
50. T. Saito, M. Rehmsmeier, Precrec: fast and accurate precision–recall and ROC curve calculations in R. *Bioinformatics* **33**, 145-147 (2017).
51. R. Kannan, V. Vasanthi, in *Soft Computing and Medical Bioinformatics*. (Springer, 2019), pp. 63-72.
52. Y. Li *et al.*, Towards high-throughput microstructure simulation in compositionally complex alloys via machine learning. *Calphad* **72**, 102231 (2021).
53. G. Bonaccorso, *Machine learning algorithms*. (Packt Publishing Ltd, Birmingham, 2017).
54. R. K. W. Marceau, L. T. Stephenson, C. R. Hutchinson, S. P. Ringer, Quantitative atom probe analysis of nanostructure containing clusters and precipitates with multiple length scales. *Ultramicroscopy* **111**, 738-742 (2011).
55. K. Jin *et al.*, Thermophysical properties of Ni-containing single-phase concentrated solid solution alloys. *Mater. Des.* **117**, 185-192 (2017).
56. P. E. Blöchl, Projector augmented-wave method. *Phys. Rev. B* **50**, 17953 (1994).
57. G. Kresse, J. Furthmüller, Efficiency of ab-initio total energy calculations for metals and semiconductors using a plane-wave basis set. *Comput. Mater. Sci.* **6**, 15-50 (1996).
58. G. Kresse, J. Furthmüller, Efficient iterative schemes for ab initio total-energy calculations using a plane-wave basis set. *Phys. Rev. B* **54**, 11169 (1996).
59. G. Kresse, D. Joubert, From ultrasoft pseudopotentials to the projector augmented-wave method. *Phys. Rev. B* **59**, 1758 (1999).
60. J. P. Perdew, K. Burke, M. Ernzerhof, Generalized gradient approximation made simple. *Phys. Rev. Lett.* **77**, 3865 (1996).
61. A. Baldereschi, Mean-value point in the Brillouin zone. *Phys. Rev. B* **7**, 5212 (1973).
62. H. J. Monkhorst, J. D. Pack, Special points for Brillouin-zone integrations. *Phys. Rev. B* **13**, 5188 (1976).
63. A. Shapeev, Accurate representation of formation energies of crystalline alloys with many components. *Comput. Mater. Sci.* **139**, 26-30 (2017).
64. I. V. Oseledets, Tensor-train decomposition. *SIAM J. Sci. Comput.* **33**, 2295-2317 (2011).
65. A. G. Khachaturyan, *Theory of structural transformations in solids*. (Courier Corporation, 2013).